\documentclass{PoS}
\usepackage{amsmath}

\title{VHE observations of the gamma-ray binary system LS 5039 with H.E.S.S.}

\ShortTitle{H.E.S.S. observations of LS 5039}

\author{\speaker{C. Mariaud}$^a$, P. Bordas$^b$, F. Aharonian$^b$ $^c$, M. Boettcher$^d$, G. Dubus$^e$, M. de Naurois$^a$, C. Romoli$^c$ and V. Zabalza$^f$ for the H.E.S.S. Collaboration\\
  \llap{$^a$}Laboratoire Leprince Ringuet Ecole Polytechnique, CNRS/IN2P3, F-91128 Palaiseau, France
  \llap{$^b$}Max Planck Institute fur Kernphysik, P.O. Box 103980, D 69029 Heidelberg, Germany\\
  \llap{$^c$} Dublin Institute for Advanced Studies, 31 Fitzwilliam Place, Dublin 2, Ireland \\
  \llap{$^d$}Centre of Space Research, North-West University, Potchefstroom 2520, South Africa\\
  \llap{$^e$}Institut de Planetologie et d'Astrophysique de Grenoble, BP 53, F-38041 Grenoble, France\\
  \llap{$^f$}Department of Physics and Astronomy, University of Leicester, University Road, Leicester, LE1 7RH, United Kingdom\\

  E-mail: \email{mariaud@llr.in2p3.fr}}

\abstract{LS 5039 is a gamma-ray binary system observed in a broad energy range, from radio to TeV energies. The binary system exhibits both flux and spectral modulation as a function of its orbital period. The X-ray and very-high-energy (VHE, E > 100 GeV) gamma-ray fluxes display a maximum/minimum at inferior/superior conjunction, with spectra becoming respectively harder/softer, a behaviour that is completely reversed in the high-energy domain (HE, 0.1 < E < 100 GeV). The HE spectrum cuts off at a few GeV, with a new hard component emerging at E > 10 GeV that is compatible with the low-energy tail of the TeV emission. The low 10 - 100 GeV flux, however, makes the HE and VHE components difficult to reconcile with a scenario including emission from only a single particle population. We report on new observations of LS 5039 conducted with the High Energy Stereoscopic System (H.E.S.S.) telescopes from 2006 to 2015. This new data set enables for an unprecedentedly-deep phase-folded coverage of the source at TeV energies, as well as an extension of the VHE spectral range down to $\sim$120 GeV, which makes LS~5039 the first gamma-ray binary system in which a  spectral overlap between satellite and ground-based gamma-ray observatories is obtained. 
}

\FullConference{The 34th International Cosmic Ray Conference,\\
		30 July- 6 August, 2015\\
		The Hague, The Netherlands}

\begin{document}

\section{Introduction}

\begin{figure}[t!]
\centering
\includegraphics[scale=.55]{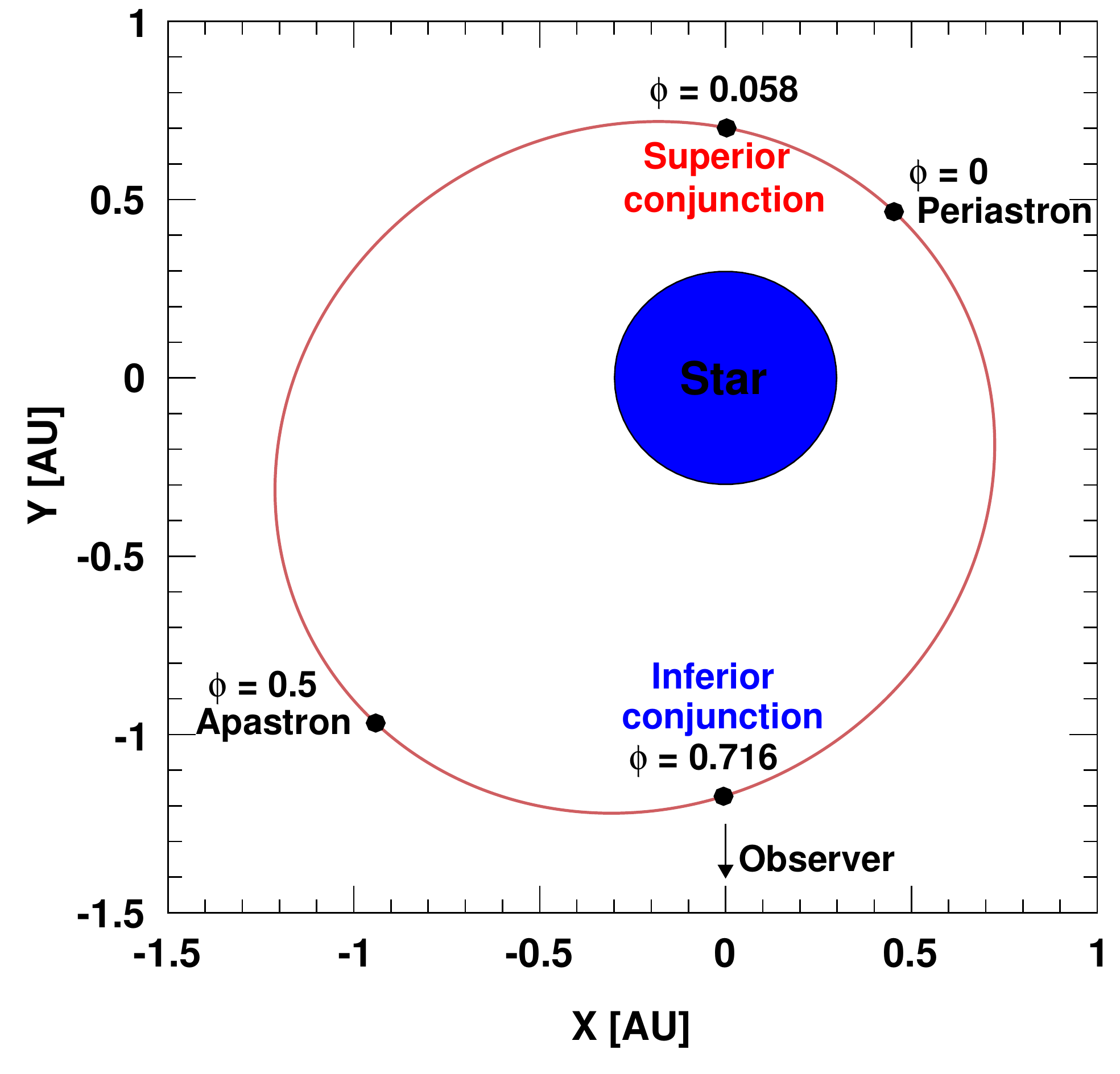}
\caption{Geometry of the orbit (in the orbital plane) of the compact object around the O6.5V star (Casares et al. \cite{2005..899..908}). The star size is to scale. The arrow at the bottom indicates the direction to the observer. Several notable positions of the compact object are indicated: periastron ($\phi = 0$), apastron ($\phi = 0.5$), superior conjunction ($\phi = 0.058$) and inferior conjunction ($\phi = 0.716$).}
\label{fig:geometry}
\end{figure}

LS 5039 is a binary system located at $\sim$~3.5 kpc from the Earth, only visible from the Southern hemisphere. It is composed of a compact object in orbit around a O6.5V star, with an orbital periodicity of $P_{\rm orb} = 3.90603$ days \cite{2005..899..908}, a moderate eccentricity ($e = 0.35$), and an angle position in the plane of the sky barely constrained ($i \approx 20 - 65^{\circ}$). The mass of the companion star is $\sim 23\; M_{\odot}$, and its radius $9.3\; R_{\odot}$ (Fig. \ref{fig:geometry}). The nature of the compact object ($M_{X} \approx 3.7\; M_{\odot}$), either a black hole or a neutron star, remains unclear.  


LS~5039 has been detected in radio, featuring persistent outflows (with sizes ranging $\sim 2-2000$ astronomical units,  AU) which classified it as a possible microquasar candidate \cite{2000..2340..2342}. In X-rays and high-energy (HE) gamma-rays, the emission appears modulated with the orbital period \cite{2009..L56..L61}. At very high energies (VHE), the High Energy Stereoscopic System (H.E.S.S.) telescopes detected the system in 2004 with $\sim$ 11 h of observations obtained during the H.E.S.S. Galactic Plane survey \cite{2005..746..749}. After further follow-up on the source, a modulation of its VHE gamma-ray flux was revealed, allowing for a determination of the periodicity in gamma-rays, with $P$ = 3.9078 $\pm$ 0.0015 d \cite{HESS_paperII}. 
%
%

Markedly different flux and spectral properties were observed close to the system's superior and inferior conjunction (SUPC and INFC, respectively). A lower gamma-ray flux and a spectrum well-fit by a simple power law (PL) is derived for SUPC. At INFC, the source is brighter and the spectrum significantly deviates from a power law above a few TeV. Both hardness and intensity at TeVs seem anti-correlated with what is obtained at HEs \cite{2009..L56..L61}. Such anti-correlation can be effectively explained by the angular dependence of both inverse Compton (IC) and pair-production cross sections and the variable observing conditions of the system along the orbit. However, the properties and location of both accelerator(s) and emitter(s) in the source are far from understood (see, e.g., \cite{Paredes2006, Dubus2006, Bosch-Ramon2008, Dubus2008, Zabalza2013, Dubus2013}, and references therein).

Since the last H.E.S.S. publication in 2006 \cite{HESS_paperII}, a substantial amount of observation time has been devoted to LS~5039. In the next section we report on an updated analysis of H.E.S.S. phase I data, whereas Sect.~\ref{sec:observations}  focuses on the first results obtained with the H.E.S.S. phase II, for which the so-called \emph{mono} analysis \cite{Holler2015} (the new 28 m telescope, CT5 hereafter, in standalone mode) has been employed. A summary and main conclusions of this study are presented in Sect.~\ref{Conclusions}. All  results shown in this proceedings, albeit thoroughly cross-checked with an independent data calibration and analysis framework, are to be considered preliminary. A more detailed report is in preparation. 

\section{Observations with H.E.S.S. phase I}
\label{HESS-I}


New data on LS~5039 have been obtained from 2006 to 2012 with the H.E.S.S. phase I telescopes. In parallel, more sophisticated analysis techniques and H.E.S.S. data-reduction software have been developed, notably improving both the background rejection power, the accuracy on the energy reconstruction of the gamma-ray showers and the angular resolution of the instrument \cite{Naurois2009, Parsons2014}. Below we report on the results of this new data set, as well as a re-analysis of pre-2006 observations with the latest analysis tools, to characterise the long-term behaviour of LS~5039 at VHEs in a time-range spanning more than 8~years.  The updated H.E.S.S. I data set amounts to about $\sim$ 104 h of live observation time, with a mean zenith angle of 19.8$^{\circ}$, yielding a detection of the source at a statistical significance of more than 56$\sigma$ and 2800 detected gamma-rays. To perform this analysis, we used a ring background subtraction \cite{Berge2007} in stereo mode.



\subsection{H.E.S.S. I: phase-folded light curve}
\label{sec:Folded_Lightcurve}

\begin{figure}[t!]
\centering
\includegraphics[width=\textwidth]{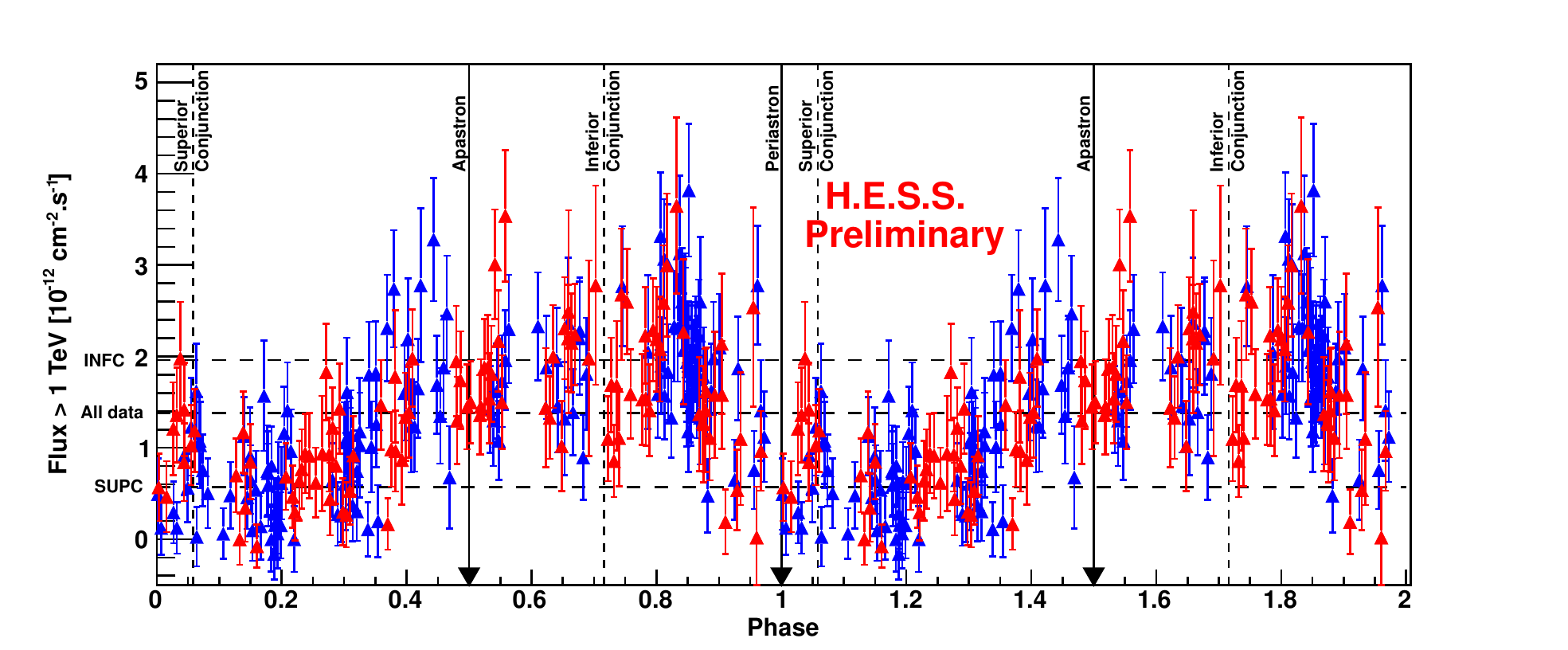}
\caption{Run-by-run light curve corresponding to pre-2006 data (\emph{red}), the same data set as reported in \cite{HESS_paperII}, reanalysed here with the same analysis tools as for the post-2006 data set (\emph{blue}). The horizontal upper, middle and lower dotted lines represent the average flux for the INFC, all-data, and  SUPC, respectively.}
\label{fig:Lightcurve_HESSI_Bis}
\end{figure}

The integral flux above 1 TeV, folded with the orbital period of the system, is computed assuming a spectral index $\Gamma = 2.20\; \pm \; 0.03$, where $\Gamma$ is derived from a fit with a simple power law, $dN/dE_{\gamma} \propto E_{\gamma}^{-\Gamma}$, to the whole data-set. In Fig.~\ref{fig:Lightcurve_HESSI_Bis} the known modulation of the source at VHE is recovered, which keeps essentially unchanged from one year to the next, for the whole data set, averaging over short, e.g., daily time-scales.

\begin{figure}[t!]
\centering
\includegraphics[width=\textwidth]{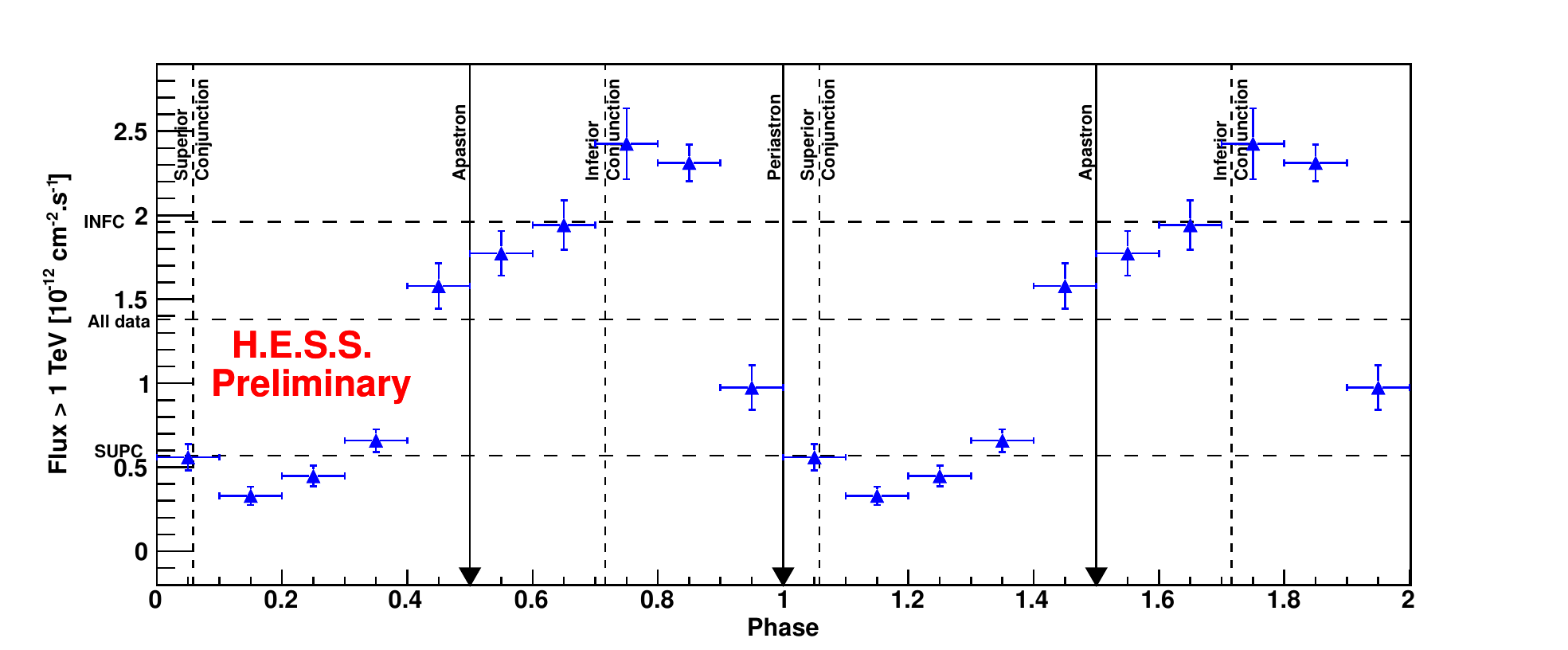}
\caption{Phase folded light curve, with LS~5039 integral flux above 1 TeV plotted over the system orbital phase. Integral fluxes are computed from a separate power-law fit in every phase interval.}
\label{fig:Average_Phase_Folded}
\end{figure}

\begin{figure}[t!]
\centering
\includegraphics[scale = 0.5, angle = 0]{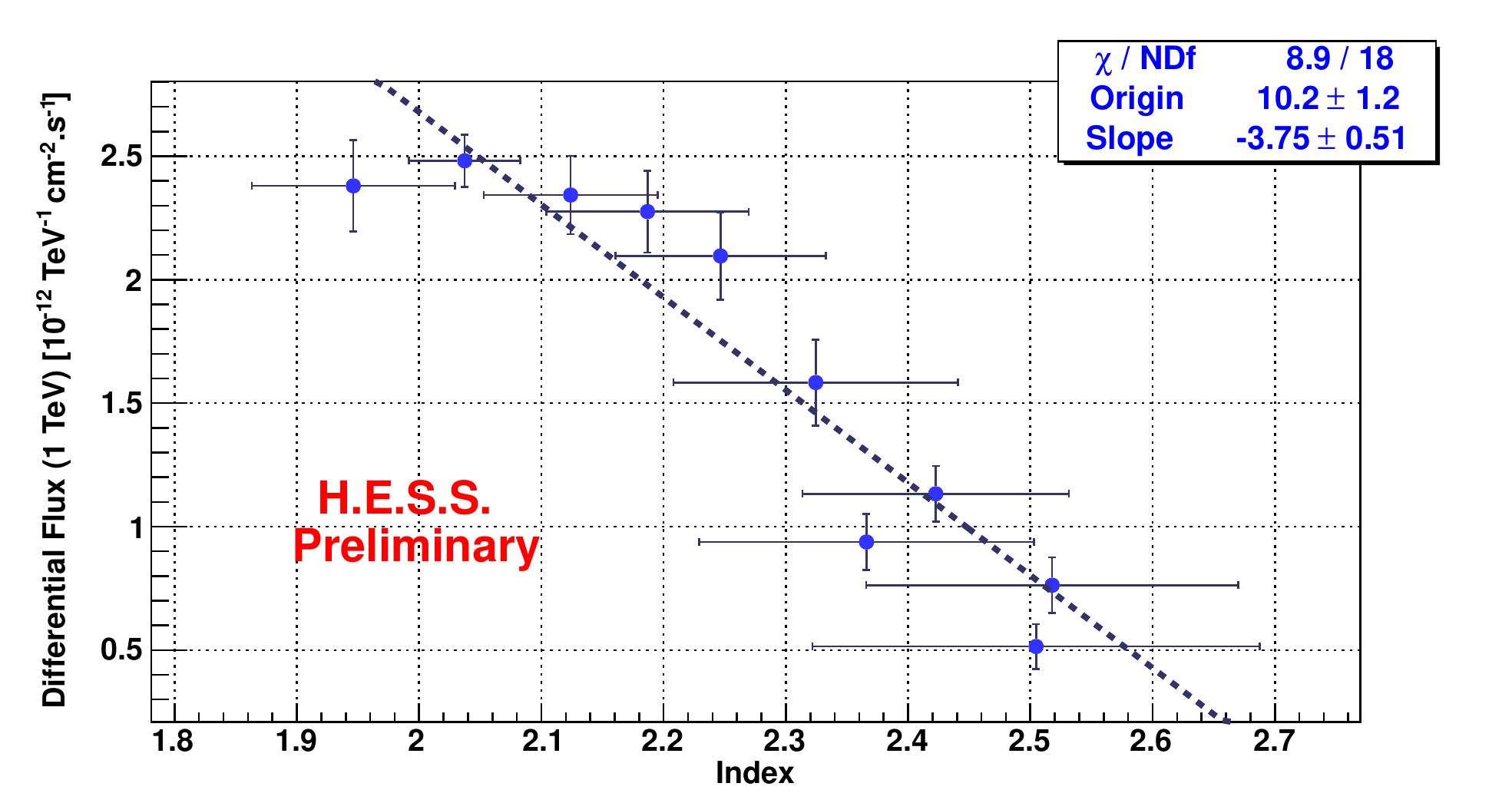}
\caption{Differential gamma-ray flux at 1 TeV versus photon index $\Gamma$ obtained from a dedicated power-law fit in each 0.1--width phase interval. The dotted line and the statistics in the legend correspond to a linear fit to the data.}
\label{fig:Norm_vs_Index}
\end{figure}

To better characterise this modulation, Fig.~\ref{fig:Average_Phase_Folded} displays the phase-folded averaged flux for phase intervals of width 0.1 each. In this case, however, the large data set obtained in the last $\sim 10$~yr of observations enables us to fit separately each phase interval to compute the bin-averaged flux. We assume here again a simple power law model fitted to the data in each phase interval. The corresponding photon index $\Gamma$ plotted against the differential gamma-ray flux at 1 TeV in the same bin is displayed in Fig.~\ref{fig:Norm_vs_Index}. We notice a relationship between both variables, the correlation coefficient is $-$0.93. When the system is brighter the spectrum is harder. In the same way, when the differential flux decreases, the spectral index increases. We note that noticeable departures from a power law may not be excluded for some phase intervals. A more detailed spectral characterisation is in preparation and will be shown in a future publication. 

The light curve in Fig.~\ref{fig:Average_Phase_Folded} displays a ratio $\gtrsim 8$ between the maximum and minimum flux levels, corresponding to the compact object being at the INFC and SUPC, respectively, with an absolute maximum at $\phi \sim 0.75 \;\pm\; 0.05 $. A detection of LS~5039 at a significance level above 5$\sigma$ is now obtained in every single phase-bin. We did not try, however, to fit the light curve neither to constrain its timing structure nor to characterise its asymmetry or the presence of one or more components, which is left for a forthcoming publication.

\subsection{H.E.S.S. I:  spectral analysis}
\label{sec:Spectrum_with_HESS_phase_I_data}

\begin{figure}[t!]
\centering
\includegraphics[scale = 0.7, angle = 0]{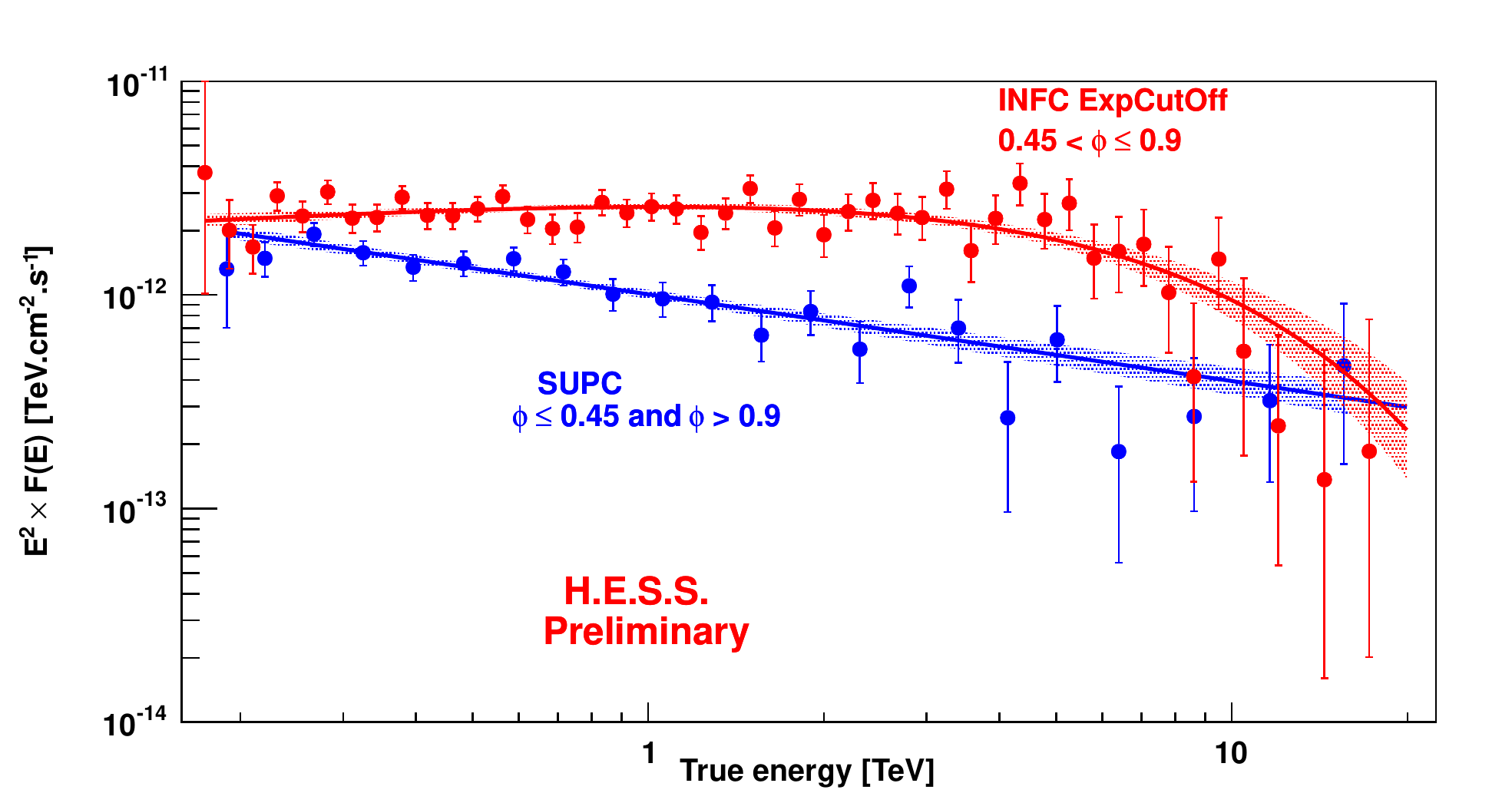}
\caption{Spectral energy distribution for the analysis of H.E.S.S. I data obtained in 2004 - 2012 The data have been divided between orbital phases close to SUPC (blue) and INFC (red), which are well fit by a simple power law and an exponential cutoff model, respectively.}
\label{fig:Spectra_INFC_and_SUPC}
\end{figure}

The large data set now available allows for a rather high-precision characterisation of the averaged spectra at both INFC (at orbital phases $\phi \in [0.45$--0.9]) and SUPC (with $\phi \leq 0.45$ or $\phi > 0.9$), see Fig \ref{fig:Spectra_INFC_and_SUPC}. A straight power law fits well the spectrum at SUPC, with an averaged spectral index of $\Gamma = 2.406\;\pm\;0.052$. 
For orbital phases close to INFC, a power law with an exponential cutoff model $\rm{d}N/\rm{d}E_{\gamma} \propto E_{\gamma}^{-\Gamma} \, \times \rm{exp} (- E_{\gamma}/E_{\rm cut})$ has been fitted (with $\Gamma = 1.843\;\pm\;0.063$, and an energy cut $E_{\rm cut} = 6.6\;\pm\;1.6 $ TeV) in agreement with previous results \cite{HESS_paperII}. Further spectral models, e.g. a broken-power law or a log-parabolic distribution are not reported here, although they also provide a better fit to the INFC data compared to a simple power law assumption. 



\section{Observations with H.E.S.S. phase II}
\label{sec:observations}

\subsection{H.E.S.S. II: dataset and analysis used}

LS~5039 has been observed during the commissioning of H.E.S.S. II in 2012-2013, with further observations taken in 2014 and 2015. Here we present the first results of these observations, which amount to a total of 39 runs taken towards LS~5039 and yielding a total observing time of 18.5 h of data after quality selection cuts. The data have been analysed in the so-called \emph{mono} mode, making use of the newly deployed CT5 telescope in standalone mode, which permits to reach the lowest energy threshold attainable with H.E.S.S. II. An analysis performed within the so-called \emph{combined} mode, with a similar performance at the lowest energies but with additional \emph{stereo}-mode sensitivity above tens of TeV \cite{Holler2015}, will be presented in a forthcoming publication. This analysis makes use of a technique based on a semi-analytical shower development model \cite{Naurois2009}, adapted to the CT5 \emph{mono} mode \cite{Holler2015} observations. All results have been cross-checked with an independent calibration software and analysis chain (ImPACT, \cite{Parsons2014}, also adapted to CT5 \emph{mono} observations), providing compatible results.

\begin{figure}[t!]
\centering
\begin{tabular}{cc}
\centering
   \includegraphics[width = 6cm, angle = 0]{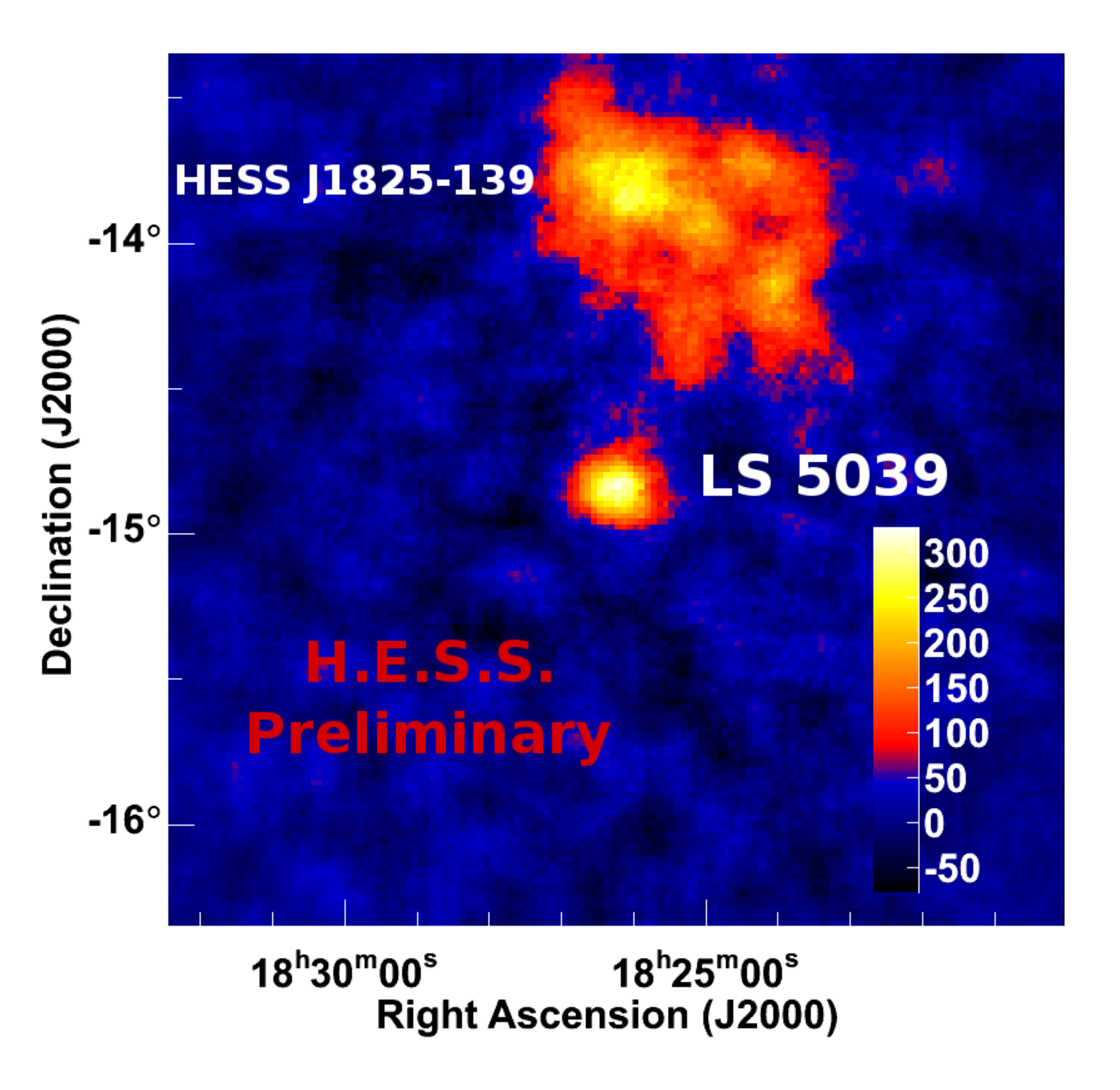} &
   \includegraphics[width = 6cm, angle = 0]{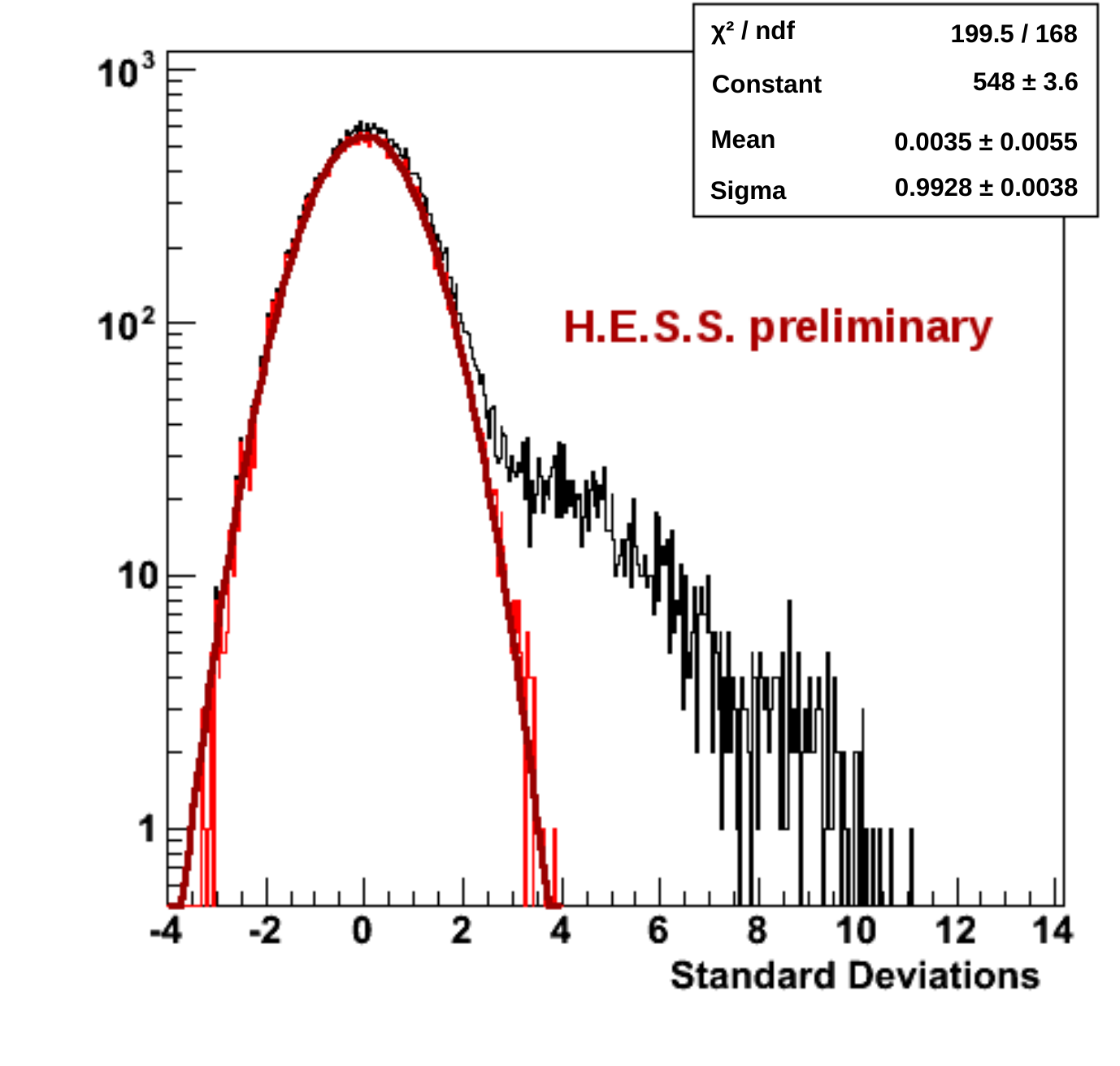} \\
\end{tabular}
\caption{LS 5039 excess map from a ring background method, for a mono analysis with $\sim$ 19 h of observation time and safe cuts, is displayed in the left panel. LS~5039 is clearly detected, together with the nearby, extended pulsar wind nebula HESS J1825--139. The distribution of the background events, after excluding both LS~5039 and HESS J1825--139, is displayed on the right panel, featuring a well-centered and well-normalised significance distribution.}
\label{excess_map}
\end{figure}

The results reported here account for observations taken under zenith angles ranging in between 9$^{\circ}$ and 52$^{\circ}$, with an average of 31$^{\circ}$. Data were taken in \emph{wobble} mode, where the telescope alternates pointing directions with a given offset from the source position. We make use of the \emph{Ring} background method \cite{Berge2007} to produce sky-maps and significance distributions in the FoV (see left panel in Fig.~\ref{excess_map}. LS~5039 is clearly detected with at a statistical significance of 12$\sigma$ (with 342 gamma-rays). In addition, the nearby source HESS~J1825--139 is also clearly detected in this analysis, demonstrating the capabilities of this CT5 \emph{mono} mode to reconstruct the signal from extended sources. The background significance distribution after subtracting both LS~5039 and HESS~J1825--139 is well normalised, as displayed in the right panel of Fig.~\ref{excess_map}. 
\\




\subsection{H.E.S.S. II: spectral analysis}

We performed a CT5 \emph{mono} spectral analysis in order to reconstruct events in the $\sim 0.1-2$ TeV energy range. The \emph{Reflected} background method, in which OFF events  are taken from several regions at the same radial distance with respect to the camera centre as the target position, is employed to retrieve LS~5039's spectrum. \emph{Standard} cuts for the background subtraction have been used, allowing for a robust characterisation of LS~5039 spectra down to a threshold energy of 119~GeV for the current data set analysis. Although lower energies may be reached after accounting for a larger H.E.S.S. data set, for which a dedicated study is in preparation, we note that this preliminary analysis already allows for the study of the energy range covered by both H.E.S.S. and the \emph{Fermi}-LAT. This is the first time that a full GeV-TeV overlap is obtained for any gamma-ray binary system.

\begin{figure}[t!]
\centering
\includegraphics[width=\textwidth]{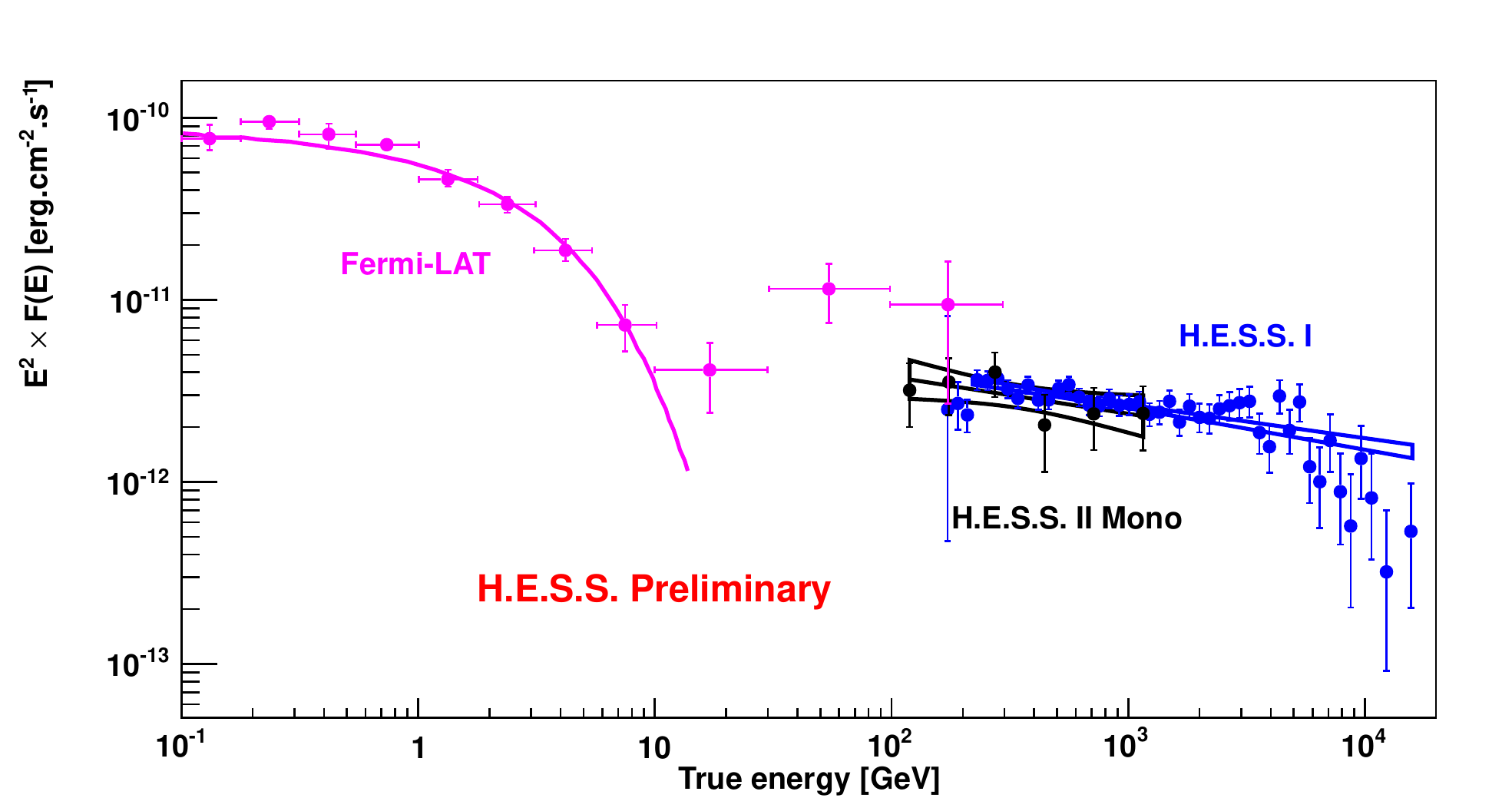}
\caption{SED of LS~5039 including the analysis of H.E.S.S. phase-II data in CT5 \emph{mono} mode (black), observations taken between 2004 and 2012 with the H.E.S.S. phase-I telescopes (blue), and \emph{Fermi}-LAT spectral points (purple) extracted from Takata et al. (2014) \cite{2014ApJ..790..18T}, with a purple solid line representing the fit with a power law with an exponential cutoff in the energy range 0.1--10 GeV (with $\Gamma = 2.06\;\pm\;0.02$ and $E_{cut}= 3.42\;\pm\;0.17$ GeV).}
\label{fig:Spectra_HESSII}
\end{figure}

A spectral energy distribution (SED) of LS~5039 at gamma-rays is displayed in Fig.~\ref{fig:Spectra_HESSII}. The spectral fit assumes a pure power-law model, from which a spectral index of $\Gamma = 2.20\; \pm \; 0.03$ is retrieved, in perfect agreement with the analysis of (non contemporaneous) H.E.S.S. phase I analysis in the overlapping energy range. The H.E.S.S. phase II CT5 \emph{mono} spectrum is compatible within the relatively large error bars of the $\gtrsim 100$~GeV flux obtained in the analysis of \emph{Fermi}-LAT data reported in \cite{2014ApJ..790..18T}. A refined analysis of the H.E.S.S. data, accounting for further observations of LS~5039 with CT5, as well as an updated \emph{Fermi}-LAT data set, will provide a deeper overlap of both instruments in a wider energy range, allowing for a more detailed study of the source spectral properties.


\section{Summary}
\label{Conclusions}

The results reported here for the analysis of H.E.S.S. phase I data, both accounting for a reanalysis of previous observations with more sensitive analysis tools and the inclusion of new data taken from 2006 to 2012, confirms the regular behaviour of LS 5039 in a time range spanning more than $\sim 8$~yr. The evolution of both the VHE gamma-ray flux level and spectral properties depending on the orbital position of the compact object in the system are confirmed with this new analysis. A more detailed study, including phase-folded spectral characterisation in relatively thin phase ranges, the characterisation of the degree of contamination of the nearby source HESS~J1825--139, and the implications in terms of emission/absorption processes responsible for the TeV emission, is in preparation. 

The analysis of the new H.E.S.S. phase II observations reported here allow for a spectral derivation down to $\sim 120$~GeVs for the data-set analysis results shown here. This is the first case for any known gamma-ray binary system in which a spectral overlap between satellite and ground-based observatories is obtained. Given the growing evidence that GeV and TeV emission are produced in different regions and/or by separate particle populations,  a deeper study of this overlapping energy range may provide strong constrains to the physics behind the gamma-ray emission in LS~5039 and possibly other similar systems.  

\vspace{0.5cm}

\noindent{\small{ACKNOWLEDGMENTS: please see standard acknowledgement in H.E.S.S. papers, not reproduced here due to lack of space.}}


\end{document}